\documentclass[aps,showpacs,preprint,superscriptaddress,nofootinbib]{revtex4}
\usepackage{graphicx}
\usepackage{subfigure}
\usepackage{float}
\usepackage{amsmath}
\usepackage{amsfonts}
\usepackage{color}
\usepackage{bm}
\usepackage{bbm}
\usepackage{txfonts}
\usepackage{array}
\usepackage{amssymb}

\begin{document}

\title{Application of partial wave analysis in multiphoton pair production}

\author{Hong-Hao Fan}
\affiliation{Key Laboratory of Beam Technology of the Ministry of Education, and School of Physics and Astronomy, Beijing Normal University, Beijing 100875, China}

\author{Li-Na Hu}
\affiliation{Key Laboratory of Beam Technology of the Ministry of Education, and School of Physics and Astronomy, Beijing Normal University, Beijing 100875, China}

\author{Suo Tang}
\affiliation{College of Physics and Optoelectronic Engineering, Ocean University of China, Qingdao, Shandong, 266100, China}
	
\author{Zi-Liang Li}
\affiliation{School of Science, China University of Mining and Technology, Beijing 100083, China}

\author{Bai-Song Xie}\email{bsxie@bnu.edu.cn}
\affiliation{Key Laboratory of Beam Technology of the Ministry of Education, and School of Physics and Astronomy, Beijing Normal University, Beijing 100875, China}
\affiliation{Institute of Radiation Technology, Beijing Academy of Science and Technology, Beijing 100875, China}

\date{\today}
\begin{abstract}
Electron-positron pair production is investigated in the multiphoton regime under polarized fields. Partial wave analysis can be applied to reveal the momentum spectral structure and identify the positions of valleys/peaks of the multiphoton rings by combining with $CP$ conservation. Moreover, considering the spin states of created pairs in partial wave analysis, the particle orbital angular momentum associated with the valleys/peaks can also be identified from multiphoton rings. It is found that the results for angular distribution by this approximate analytical treatment and by the Dirac-Heisenberg-Wigner formalism are consistent with each other. The present results are helpful for understanding the angular momentum information in momentum spectra, and they also have an implication for revealing the vortex structure near the valleys.
\end{abstract}
\maketitle

\section{Introduction}

The production of electron-positron pairs from vacuum~\cite{Heisenberg:1936nmg,Schwinger:1951nm,Sauter:1931zz} is a significant theoretical prediction of quantum electrodynamics (QED). Due to the current electric field strengths being far below the critical electric field strengths $E_{\text{cr}} \sim 10^{16} $ V/cm, detecting vacuum pair creation experimentally remains challenging. However, the high-intensity laser pulse facilities under construction~\cite{Xfel:1111link,VBagnoud:2009,Hipper:1111link,ELI:2006eli}
will provide subcritical electric fields for pair production in the near future~\cite{Dittrich:2000w,Ringwald:2001,Alkofer:2001ik}. In 1997, the Stanford Linear Accelerator Center (SLAC) E-144 experiment successfully observed the production of electron-positron pairs through the absorption of $4$-$5$ photons during high-energy electron-laser collisions~\cite{SlacE144:1997}. This mechanism offers more experimental opportunities for vacuum pair production via multiphoton absorption, sparking renewed interest in this research area. In past decades, various fields configuration have been employed to study multiphoton pair production~\cite{PhysRevD.2.1191,Ritus:1985vi,Baier:1998vn,RalfSchtzhold:2008hg,Wollert:2014epy,Xie:2017xoj,Aleksandrov:2018uqb,Huang:2019uhf,Kohlfurst:2019mag,Mohamedsedik:2021M}. These studies have revealed significant effects, including the ponderomotive effect~\cite{Kohlfurst:2018c}, effective mass characteristics~\cite{Kohlfurst:2013ura}, and spiral structures~\cite{Ngokoz:2015,Li:2017qwd,Hu:2023pmz}.

In the multiphoton process, the pairs absorbing energy from photons often exceed the minimum required energy, resulting in a series of valleys/peaks in the momentum spectra~\cite{Li_2015,Kohlfurst:2019mag}.
The distribution of valleys/peaks in the momentum spectra is related to the number of absorbed photons and the frequency of the external fields.
By using partial wave analysis (PWA), the positions of valleys/peaks can be precisely determined,
and the contributions of different spin states in momentum spectra can also be extracted under circularly polarized fields~\cite{Kohlfurst:2018kxg,Yang19480nTA}.
Although such operation is performed in circularly polarized fields, this method is adaptable to other field configurations.
Moreover, under linearly and elliptically polarized cases, the valley/peak structures on the multiphoton rings are more complex.
The contributions of the spin states and the angular momentum distribution at these positions require further confirmation and detailed investigation.

In this paper, we use the real-time Dirac-Heisenberg-Wigner (DHW) formalism~\cite{PhysRevD.44.1825,Blinne:2013via,Weickgenannt:2019dks,Sheng:2018jwf,Mameda:2023ueq,Hebenstreit:2011fh,Ababekri:2019dkl,Aleksandrov:2019ddt,Ababekri:2019qiw,Li:2021wag} to obtain the phase space distribution function of the created particles in the time-dependent external fields with arbitrary polarization.
We then investigate the angular distribution of the momentum spectra by using PWA.
Additionally, we also take into account charge ($C$) conjugation and parity ($P$) conservation as well as angular momentum conservation.
Based on $CP$ conservation, we provide a semi-quantitative description of the valleys/peaks in multiphoton rings under arbitrarily polarized fields.
Under angular momentum conservation, we extract the different spin-state probabilities from the total probability and obtain the orbital angular momentum (OAM) of the created particle.

Note that we set $\hbar = c = 1$ throughout this paper, and all quantities, such as momentum and field frequency, are given in terms of electron mass $m$.

This paper is organized as follows. In Sec.~\ref{SecII}, we give the external fields configuration and the DHW formalism.
In Sec.~\ref{SecIII}, we present the particle distribution of multiphoton absorption in arbitrarily polarized fields.
In Sec.~\ref{SecV:CP}, we determine the positions of the valleys/peaks based on $CP$ conservation.
In Sec.~\ref{SecIV:AD}, we provide the angular momentum information in the momentum spectra.
In Sec.~\ref{summary}, we give a summary.

\section{External field model and theoretical formalism}\label{SecII}

We consider a toy model of external fields
\begin{equation}\label{Eq:1}
\begin{aligned}
\boldsymbol{A}(t)
&= \frac{E_0}{\omega\sqrt{1+\delta^2}}\exp(-\frac{t^2}{\tau^2})
\begin{pmatrix}
  \sin(\omega t) \\
  \delta \cos(\omega t) \\
  0
\end{pmatrix},
\end{aligned}
\end{equation}
where $E_0$ is the field strength, $\omega$ is the frequency, $\delta$ denotes the polarization of the field, and $\tau$ is the pulse duration.
Throughout this paper, the fixed field parameters are chosen as: $E_0 = 0.2E_\text{cr}$, $\omega = 0.6m$, and $\tau = 25m$.
Meanwhile, we consider three typical polarization parameters: $\delta = 0$, $0.5$, and $1$.

These configurations represent left-handed and right-handed waves counterpropagating to form a standing wave.
For linearly polarized fields $\delta = 0$, the helicity in the standing wave is determined by $n_+$ photons corresponding propagating in the $+z$ direction and $n_-$ photons propagating in the $-z$ direction~\cite{Ruf:2008ahs,DiPiazza:2011tq}.
The net photon absorption number $l_N = n_+ - n_-$ will give a total helicity of $l_N$~\cite{Aleksandrov:2019ddt}.
Without loss of generality, every photon in the lasers propagates along the $+z$ direction for the helicity $h = 1$ under the circularly polarized fields $\delta = 1$.
However, for elliptically polarized fields with $0 < \delta < 1$, it is quite difficult to provide a precise characterization of its helicity.

The widely adopted DHW formalism in strong background fields is shown in Refs.~\cite{Hebenstreit:2011pm,kohlfurst2015electronpositron,Sheng:2019ujr}.
For time-dependent external fields, the single-particle distribution function $f(\boldsymbol{p},t)$ can be obtained by solving the following ordinary differential equations~\cite{Blinne:2016yzv}
\begin{equation}\label{Eq2:ODEs}
  \begin{array}{l}
    \dot{f}=\frac{ e\bold{E} \cdot \boldsymbol{{v}}}{2 \omega_p}, \\
    \dot{\boldsymbol{v}}=\frac{2}{\omega_p^{3}}\left[( e\bold{E} \cdot \boldsymbol{p}) \boldsymbol{p}- e\bold{E}\omega_p^{2} \right]
    (f-1)-\frac{( e\bold{E} \cdot \bold{v}) \boldsymbol{p}}{\omega_p^{2}}-2 \boldsymbol{p} \times \mathbbm{a}-2 m \mathbbm{t}, \\
    \dot{\mathbbm{a}}=-2 \boldsymbol{p} \times \boldsymbol{v}, \\
    \dot{\mathbbm{t}}=\frac{2}{m}\left[m^{2} \bold{v}+(\boldsymbol{p} \cdot \boldsymbol{v}) \boldsymbol{p}\right],
  \end{array}
\end{equation}
where the dot represents the derivative with respect to time.
The initial conditions are $f(\boldsymbol{p},-\infty) = \boldsymbol{v}(\boldsymbol{p},-\infty) = \mathbbm{a}(\boldsymbol{p},-\infty) = \mathbbm{t}(\boldsymbol{p},-\infty) = 0$.
The single-particle energy is $\omega_{\boldsymbol{p}} = \sqrt{m^2 + \boldsymbol{p}^2}$, where $\boldsymbol{p} = \boldsymbol{q}-e\boldsymbol{A}(t)$ is kinetic momentum which can be considered as canonical momentum $\boldsymbol{q}$ for $\boldsymbol{A}(t \to \pm \infty) = 0$ at asymptotic time.

\section{Momentum spectrum and angular distribution}\label{SecIII}

In ($p_x,p_y$) planes, the momentum spectra are axially symmetric with respect to $p_x = 0$.
For simplicity, we set $p_y = 0$ to investigate the structures of momentum spectra in the ($p_x$, $p_z$) planes.
For the $n$-photon process, the created particles also receive contributions from the production process involving $n = n_{min} + s$ photons for $s = 0, 1, 2,\ldots$, where $n_{min} = \lceil\frac{2m}{\omega}\rceil= 4$ and the symbol $\lceil \cdot \rceil$ denotes the ceiling function.

\begin{figure}[htbp]
  \includegraphics[scale=0.3]{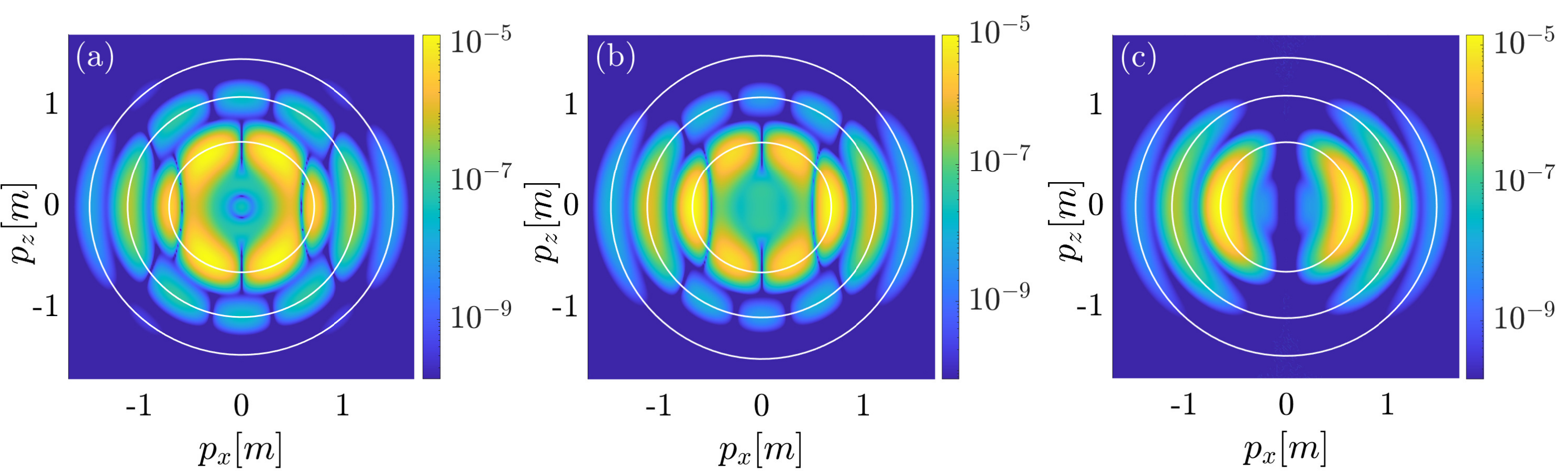}
  \caption{Particle momentum spectra $f(p_x, p_z)$ for $p_y = 0$. The white curves are determined by the $4$-, $5$- and $6$-photon channel.
  From left to right correspond to $\delta = 0$, $0.5$ and $1$ respectively.
  External field parameters are $E_0 = 0.2E_{cr}$, $\omega = 0.6m$ and $\tau = 25m^{-1}$}
  \label{Fig:1}
\end{figure}

From Fig.~\ref{Fig:1}, we can observe the pronounced features of multiphoton absorption.
The white curves in the momentum spectra in Fig.~\ref{Fig:1}, from inner to outer, represent the $4$-, $5$- and $6$-photon channels corresponding to $s = 0$, $1$ and $2$, respectively.
We can understand the structures of the distribution function in phase space from the perspective of energy conservation.
The effective energy $\mathcal{E} \left(\boldsymbol{p}\right)$ of the created particles is given by~\cite{Otto:2014ssa,Aleksandrov:2018uqb}:
\begin{equation}\label{Eq:3}
\mathcal{E}\left(\boldsymbol{p}\right) = \frac{1}{2\pi} \int^{2\pi}_0 dt \sqrt{m^2 + \left[\boldsymbol{p} - e \boldsymbol{A}(t)\right]^2},
\end{equation}
and it satisfies the relation $2 \mathcal{E}\left(\boldsymbol{p}\right) = n \omega$.
The Eq.~\eqref{Eq:3}, where the temporal envelope in the vector potential $\boldsymbol{A}(t)$ has been stripped out for simplicity, demonstrates that the shape of the $n$-photon ring is determined by a fixed energy $\mathcal{E}\left(\boldsymbol{p}\right)$.
In a linearly polarized field, the particle is strongly accelerated in the $x$ direction, leading to a large distribution in this direction, so elliptic shapes for the $n$-photon channel appear, see the white curves as shown in Fig.~\ref{Fig:1}(a). Similarly, for $\delta = 0.5$, the shapes in the momentum spectra approach a circle as shown in Fig.~\ref{Fig:1}(b), and form a circle in the case of $\delta = 1$ in Fig.~\ref{Fig:1}(c).
From Fig.~\ref{Fig:1}, we can also observe that the momentum spectra are symmetric for both $p_x = 0$ and $p_z = 0$ axes.
This is because the integrand in Eq.~\eqref{Eq:3} remains constant under the transformation $\omega_{\boldsymbol{p}} (p_x, p_y, p_z, t) \to \omega_{\boldsymbol{p} }(-p_x, p_y, -p_z, -t)$.

\begin{figure}[htbp]
  \includegraphics[scale=0.33]{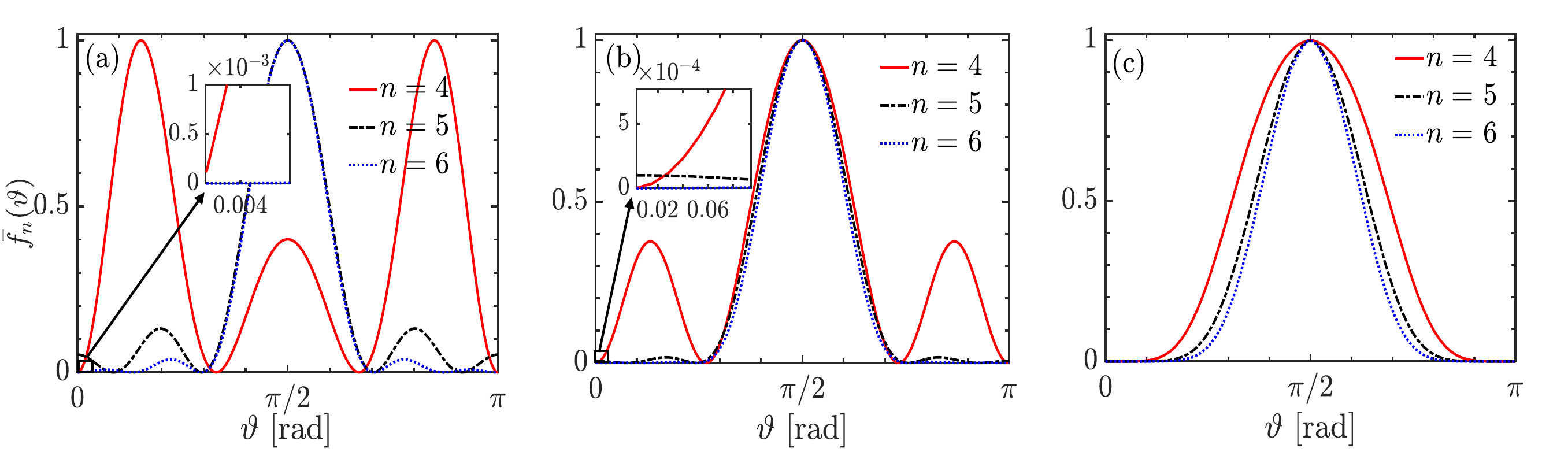}
  \caption{(color online). Normalized angular distribution function of $n$-photon absorption for $\delta = 0$, $ 0.5$ and $1$ correspond to (a), (b) and (c).
  The angle $\vartheta$ ranges  from $0$ to $\pi$ in a clockwise direction, starting from $p_x=0$.
  Other parameters are same as in Fig.~\ref{Fig:1}.}
   \label{Fig:2}
\end{figure}

It is clear that the momentum spectra exhibit complex valley/peak structures in the multiphoton rings.
In order to identify the valley/peak positions more easily, we present the angular distributions for different photon number absorptions in Fig.~\ref{Fig:2}.
By interpolating along the white curves, we can present the distribution function as a function of the angle $\vartheta$ for $n$-photon channels, i.e.,  $f_n(\vartheta) = f_n(p_r,\vartheta)$.
Here, $p_r$ is defined as $p_r = \sqrt{p_x^2+p_z^2}$ and $\vartheta$ as the angle between the particles' ejected direction and the $z$ direction.
Due to symmetry, we can derive that $f(-p_x,p_z) = f(p_x,p_z)$, so we only consider the positive half-plane where $p_x > 0$.

For convenience, we normalize the relative intensity results to $ \bar f_n(\vartheta) = \frac{f_n(\vartheta)}{I_n^{\text{max}}} $, where $I_n^{\text{max}}$ represents the maximum intensity of the $n$-photon ring.
We focus on the positions of valleys/peaks for $\vartheta \leq \pi/2$ due to the distribution symmetry around $\vartheta = \pi/2$.
We use the typical $4$- and $5$-photon absorption as examples.
When $\delta = 0$, the red curve in Fig.~\ref{Fig:2}(a) for 4-photon absorption shows that the valleys are localized at $\vartheta \approx 0$ and $0.34\pi$, and the peaks are at $\vartheta \approx 0.15\pi$ and $\vartheta = \pi/2$.
When $\delta =0.5$ the red curve in Fig.~\ref{Fig:2}(b) shows the valleys localized at $\vartheta \approx 0$ and $0.28\pi$, and the peaks are at $\vartheta \approx 0.11\pi$ and $\vartheta = \pi/2$.
In the process of $5$-photon absorption with $\delta = 0$, the valleys occur around $0.08\pi$ and $0.31\pi$, while the peaks are found at $\vartheta \approx 0$, $0.2\pi$  and $\pi/2$, see the black curve in Fig.~\ref{Fig:2}(a).
For $\delta = 0.5$, see the black curve in Fig.~\ref{Fig:2}(b),  the valleys occur around $0.07\pi$ and $0.25\pi$, while the peaks are found at $\vartheta \approx 0$, $0.17\pi$ and $\pi/2$.
For $\delta = 1$, the minimum valley is located at $0$ and maximum peak of the created particles occur at $\pi/2$, see Fig.~\ref{Fig:2}(c).

The structure of valleys/peaks in the rings associated with the number of absorbed photons is closely related to the angular momentum of the created particles.
In the following investigation, we will discuss the angular momentum information in momentum spectra under arbitrarily polarized fields.

\section{{$CP$} conservation}\label{SecV:CP}

In this section, we use PWA to describe the valley/peak structures in Fig.~\ref{Fig:2}.
For our field models, photons do not carry OAM. Instead, their total angular momentum (TAM), denoted by $J^{\text{photon}}$, is equal to their spin angular momentum, $S^{\text{photon}}$.
The spin of the pair can couple in two ways: spin parallel ($S = 1$) and spin antiparallel ($S = 0$).
The TAM of the pair originating from the photons can be determined from the coupling of the spin angular momentum and OAM, namely, $J^{\text{photon}} = J^\text{e} = L + S$, where $J^\text{e}$ and $L$ denote the TAM and OAM of the created particles.

Performing PWA for an $n$-photon process, the final-state wave function can be expressed
\begin{equation}\label{Eq:4}
  \psi_n(\vartheta, \phi) = \sum_{L = 0}^{n}\sum_{M=-L}^{L} C_{L, M} Y_{L, M}(\vartheta, \phi), \quad(\vartheta \in[0, \pi], \phi \in[0,2 \pi]),
\end{equation}
where $C_{L, M} =\int Y_{L, M}^{*}(\vartheta,\phi) \psi_n(\vartheta,\phi) \mathrm{d} \Omega$ are the expansion coefficients, and $M$ is the magnetic quantum number. 
Considering the momentum distribution function in the $(p_x, p_z)$ planes requires $\phi = 0$, and $Y_{L, -M} = (-1)^M Y_{L, M}$, indicating that these states can be seen as the same quantum state.
Meanwhile, the odd powers of the cosine function in the summation Eq.~\eqref{Eq:4} are omitted automatically due to the symmetry requirement $\bar f_n(\vartheta) = \bar f_n(\pi - \vartheta)$.
Ultimately, the angular distribution can be expressed as
\begin{equation}\label{Eq:5}
  \bar f_n(\vartheta) = \vert \psi^{S = 0}_n(\vartheta)\vert^2 + \vert\psi^{S = 1}_n(\vartheta)\vert^2,
\end{equation}
for $n$-photon absorption based on angular momentum conservation.

In a pure neutral system, the $C$ for the photons and pair are $(-1)^{n}$ and $(-1)^{L+S}$~\cite{Griffiths:2008zz}. If the pairs are $S = 1$, even $n$ gives odd $L$, otherwise $n$ and $L$ are of the same parity with $S = 0$.
The $P$ for particle-antiparticle pair with spin-$\frac{1}{2}$ is $(-1)^{L+1}$. For $n$ photons, we can write $P$ as $(-1)^n$ under electric dipole transition~\cite{Kohlfurst:2019mag}.
Due to $C$ and $P$ being conserved quantities, and $CP$ conservation requiring $S = 1$, this indicates that the consideration of $P$ can lead to the loss of information for $S = 0$ under electric dipole transition.

We use $4$- and $5$-photon channels as examples.
The former case restricts the OAM of the created particles to be odd, while the latter requires it to be even under $CP$ conservation.
Extracting the $S=1$ contributions in Eq.~\eqref{Eq:5}, we have $\bar f_n(\vartheta) \approx \vert \psi^{S = 1}_n(\vartheta)\vert^2$.
It is not difficult to verify that the angular distribution can be written as $\bar f_4(\vartheta) \approx \vert b_1\text{sin}\vartheta +  b_2 \text{sin}^3\vartheta \vert^2$ and $\bar f_5(\vartheta) \approx \vert c_1 + c_2\text{sin}^2\vartheta + c_3 \text{sin}^4\vartheta \vert^2$ for the cases of $\delta = 0$ and $\delta = 0.5$.
Taking the first derivative with respect to $\vartheta$,
$\frac{\partial \bar f_{4/5}}{\partial \vartheta} = \frac{\partial \bar f_{4/5}}{\partial \text{sin}\vartheta }\text{cos}\vartheta = 0$, the extreme points can be identified at $\vartheta = 0$ and $\pi/2$.
The former position corresponds to the valley for the $4$-photon process and peak for the $5$-photon process.
The expansion coefficients can be easily calculated as $b_1 \approx -2.854$, $b_2 \approx 3.445$, $c_1 \approx -0.219556$, $c_2 \approx -0.731852$, and $c_3 \approx 1.951605$ for $\delta = 0$.
For $\delta = 0.5$, $b_1 \approx 1.785$, $b_2 \approx -2.8035$, and $c_i$ for $i = 1,2,3$ remain unchanged.

\begin{table}[htbp]
  \caption{The intensity of valleys/peaks are given by DHW (the third column) and predicted by $CP$ conservation (the fourth column) in non-circularly polarized fields for $n = 4$.
  Due to symmetry, we only provide the positions of valleys/peaks for $\vartheta \leq \pi/2$, see Fig.~\ref{Fig:2}.}
  \label{TableI:4photon}
  \centering
  \begin{tabular}{p{1cm}p{4cm}<{\centering}p{6cm}<{\centering}p{1.6cm}<{\centering}}
    \hline
      \hline
      $\delta$ & positions & $\bar f_4(\vartheta)$~(DHW)& $\bar f_4(\vartheta)~(CP)$ \\
      \hline
       $0$        & $0$                 & $ 2.2\times 10^{-6}$                    & $0$ \\
       $ 0$       &$0.15\pi$          & $ 1$                                       &$ 0.95$ \\
       $0$       &$0.34\pi$           &$6.5\times 10^{-5}$                         &$ 3.3 \times 10^{-2} $ \\
       $ 0$       &$\pi/2$            &$0.40$                                    &$0.35$ \\
       $0.5$      & $0$               &$ 5.4\times 10^{-5}$                        &$0$ \\
       $0.5$      &$0.11\pi$          &$0.38$                                     &$ 0.29$ \\
       $0.5$      &$0.28\pi$         &$3.9\times 10^{-4}$                       &$ 8.6 \times 10^{-3}$ \\
       $0.5$      &$\pi/2$             &$1$                                       &$1.03$ \\
      \hline
      \hline
  \end{tabular}
\end{table}

Firstly, we investigate the $4$-photon absorption in non-circular polarization.
For $\delta = 0$, substituting these positions of valleys and peaks into the partial wave expansion, we obtain $\bar{f}_4(0) \approx 0$, $\bar{f}_4(0.15\pi) \approx 0.95$, $\bar{f}_4(0.33\pi) \approx 3.3 \times 10^{-2}$, and $\bar{f}_4(\pi/2) \approx 0.35$.
For elliptical polarization $\delta = 0.5$, the corresponding intensity values are $\bar{f}_4(0) \approx 0$, $\bar{f}_4(0.11\pi) \approx 0.25$, $\bar{f}_4(0.28\pi) \approx 8.6\times 10^{-3}$, and $\bar{f}_4(\pi/2) \approx 1.03$.
When $\bar f_4(\vartheta) \ll 1$, we can consider these $\vartheta$ as the position of valleys and vice versa for peaks.
Comparing these results with the red lines in Fig.~\ref{Fig:3}(a) and (b), it becomes evident that under the $CP$ constraint, considering only $S=1$ can predict the peak and valley positions in the $4$-photon absorption ring.
This shows that the spin parallel contribution dominates compared to the spin antiparallel contribution.
In order to clearly see the difference between the valleys/peaks intensities obtained from DHW and $CP$ conservation, we list the two results in Table~\ref{TableI:4photon}.

Secondly, for the $5$-photon absorption, by substituting $c_i$ into the partial wave expansions, the positions of valleys and peaks near $\pi/2$ can be well identified, see Table \ref{TableII:5photon}.
However, when $0 < \vartheta < 0.25\pi$, the PWA fails to predict the positions of the valleys or peaks.
This discrepancy arises because we did not account for the contribution of $S=0$, which significantly impacts the $5$-photon absorption process. 
Consequently, considering only $S=1$ cannot fully characterize all the features of $5$-photon absorption under electric dipole transition.

Particularly, in the case of circular polarization  $\delta = 1$, as shown in Fig.~\ref{Fig:3}(c),  we can offer a concise description of the angular distribution.
Due to each of the absorbed photons for $\delta = 1$ having a helicity of $1$, the angular momentum quantum number and the magnetic quantum number satisfy $L = M = n-1$.
This indicates that only $Y_{L, L}$ partial wave is retained, and therefore the angular distribution of the $n$-photon process therefore can be approximately written
\begin{equation}\label{Eq:6}
  \bar f_n (\vartheta) \approx \text{sin}^{2(n-1)}\vartheta.
\end{equation}
Eq.~\eqref{Eq:6} describes the maximum distribution of the final-state particle at $\vartheta = \pi/2$, and it indicates that the particle cannot be ejected parallel to the direction of laser propagation.

\begin{table}[htbp]
  \caption{Same as in Table \ref{TableI:4photon} but for $n = 5$. The results is presented not only for a few positions closed to the symmetric axis, but also the intensity at $\vartheta = 0$ for $\delta = 0$.}
  \label{TableII:5photon}
  \centering
  \begin{tabular}{p{1cm}p{4cm}<{\centering}p{6cm}<{\centering}p{1.6cm}<{\centering}}
    \hline
      \hline
      $\delta$ & positions & $\bar f_5(\vartheta)$~(DHW)& $\bar f_5(\vartheta)~(CP)$ \\
      \hline
       $0$       &  $0$                & $4.8\times 10^{-2}$                   &  $4.8\times 10^{-2}$ \\
       $ 0$      & $0.31\pi$            &$3.6\times 10^{-4}$                    &  $2.3\times 10^{-2}$ \\
       $0$      & $\pi/2$              & $ 1$                                   &  $1$ \\
       $0.5$   & $0.25\pi$             & $2.1\times 10^{-3}$                    &  $2.2\times 10^{-3}$ \\
       $0.5$   & $\pi/2$               & $1$                                    &  $1$ \\
      \hline
      \hline
  \end{tabular}
\end{table}

From the shape of the curves for normalized angular distribution, we find that the absorption of different photon numbers under circular polarization fields exhibits a unique scaling characteristic.
Additionally, in our absorption model, the highest power in the PWA does not exceed $2 (n-1)$, a necessary condition for the conservation of angular momentum.

The above analyses show that these valleys can be approximately determined by considering only the particle OAM for fixed $S = 1$ under $CP$ conservation.
This likely indicates that the created particles have vortex characteristics~\cite{Bechler:2023kjx,Majczak:2024hmt}.
The probability amplitude of the wave function vanishes near the valleys, potentially forming a vortex center at these points~\cite{BialynickiBirula1992TheoryOQ,Taylor2016AnalysisOQ}.
When the phase of the wave function forms a closed loop around the vortex center~\cite{Geng:2021ezr,Majczak:2022xlv}, the phase changes periodically by integer multiples of $2\pi$.

\section{Angular momentum information in momentum spectra}\label{SecIV:AD}

Under the requirement of $CP$ conservation, the partial wave analysis does not provide an accurate description of the positions of valleys/peaks of the created particles.
This may be due to $P$ conservation in the scenario of an electric dipole transition leading to some degree of symmetry breaking.
In this section, we will discard the $P$ conservation to recover the full angular momentum information in the $5$-photon process to further present the contributions of different spin states, and the OAM of created particles.

\subsection{Spin states}

\begin{figure}[htbp]
  \includegraphics[scale=0.36]{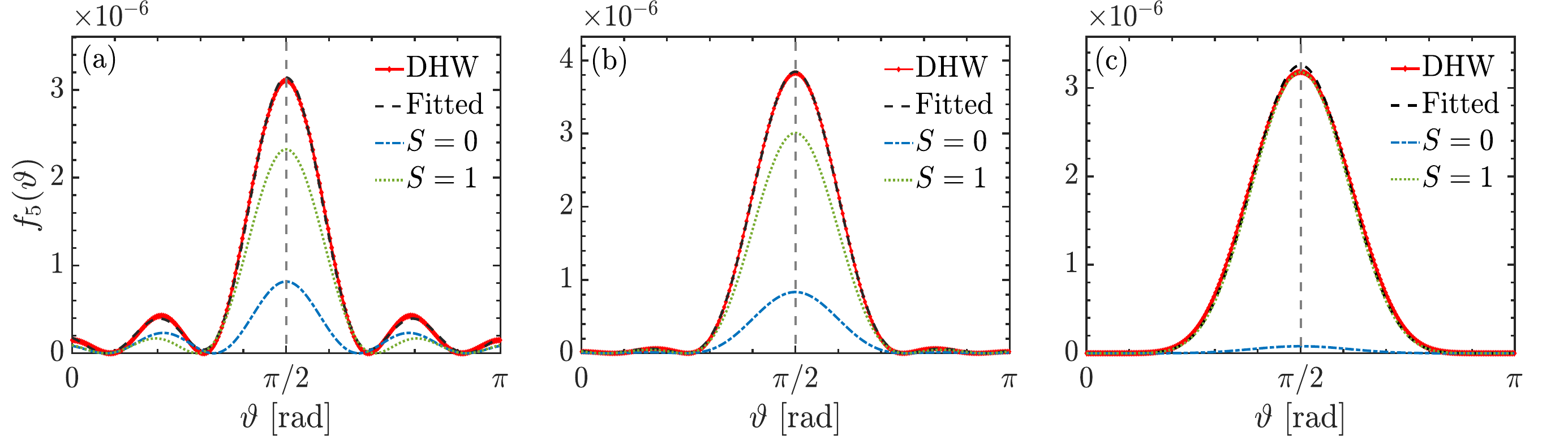}
  \caption{(color online). The contributions of different spin alignments in pair production for $5$-photon absorption.
  Angular momentum conservation distinguishes between $S = 0$ and $S = 1$ components and $\delta = 0$, $0.5$ and $1$ are labeled (a), (b) and (c).
  The parameters are same as in Fig.~\ref{Fig:1}.}
  \label{Fig:3}
\end{figure}

When performing PWA in the $5$-photon process, we choose the initial value of the coefficients as $0.2$ to start the iteration.
The probabilities of different spin states can be extracted from the total probability for $\delta = 0$, $0.5$, and $1$, as shown in Fig.~\ref{Fig:3}.
The red line in Fig.~\ref{Fig:3} corresponds to the $5$-photon angular distribution obtained by DHW, while the black line represents the result of fitting the angular distribution
using PWA. The blue and green lines represent the probabilities of $S = 0$ and $S = 1$, respectively.
The vertical gray line at $\vartheta = \pi/2$ denotes the symmetry axis.

From Fig.~\ref{Fig:3}(a)-(b) we can see that the fitting results are in perfect agreement with those obtained by DHW for $\delta = 0$, $0.5$, and $1$.
We also find that the closer $\vartheta = \pi/2$ is, the larger the spin-parallel and antiparallel contributions are. The farther from the $\vartheta = \pi/2$, the closer the spin-parallel and antiparallel contributions are.
More specifically, when $\delta = 0$, the spin antiparallel contribution is higher than the spin-parallel contribution at angles ranging from $0.25\pi$ to $0.4\pi$, as shown in Fig.~\ref{Fig:3}(a).
When $\delta = 0.5$, the spin-parallel and spin-antiparallel contributions are almost the same in the angular ranges of $0.05\pi$ to $0.11\pi$ and of $0.2\pi$ to $0.28\pi$, see Fig.~\ref{Fig:3}(b).
This explains why we cannot describe the valleys or peaks away from the symmetry axis under $CP$ conservation for $\delta = 0.5$.

Under circularly polarized fields, the spin antiparallel contribution is smaller than the spin parallel contribution, see Fig.~\ref{Fig:4}(c).
In particular, the contribution of $S = 0$ is about 30 times lower than that of $S = 1$ at $\pi/2$.

Interestingly, the created particles angular momentum pattern and the photon absorption at the different valleys/peaks differ for $\delta = 0$ and $0.5$.
Let us take the linearly polarized case as an example and consider the created particles ejected at $p_x = 0$ and $p_z = 0$.
As shown in Fig.~\ref{Fig:1}(a), the angular distribution is always a valley in the laser propagating direction for even-number photon absorptions, but a peak for odd-number photon absorptions.

Along the laser propagation direction with $p_x = 0$, the orbital magnetic quantum number is $M=0$.
The $CP$ conservation leads to a vanishing contribution of $S=1$ for even-number photon absorption.
Therefore, only the contribution of $S = 0$ exists at $p_x = 0$, and the created particles absorb an equal number of left-handed and right-handed photons, i.e., $l_N = 0$.
In the process of odd photon absorption, when electrons are ejected along the direction of laser propagation, both contribution of spin states $S = 0$ and $S = 1$ will be present. This is because the net photon numbers $\lvert l_N\rvert = 1$.
The remaining angular momentum inherited from the net photons can only be transferred to the spin of the pairs to satisfy angular momentum conservation.

When the particle is ejected perpendicular to the laser propagation direction at $p_z = 0$, the contribution of the larger OAM component wave increases. The created particles absorb non-zero net photon number resulting in the projection of TAM in the $z$ direction $J^{e}_z\neq 0$.
At this point, the spin $z$ component of the pairs satisfies $S_z = J^{e}_z - M$. If $S_z = \pm 1$ occur, a spin-parallel contribution arises.
For example, $J^{e}_z = 3$ and $M = 2$ give $S_z = 1$ and the like.

\subsection{Orbital angular momentum of the created particle}

\begin{figure}[htbp]
  \includegraphics[scale=0.5]{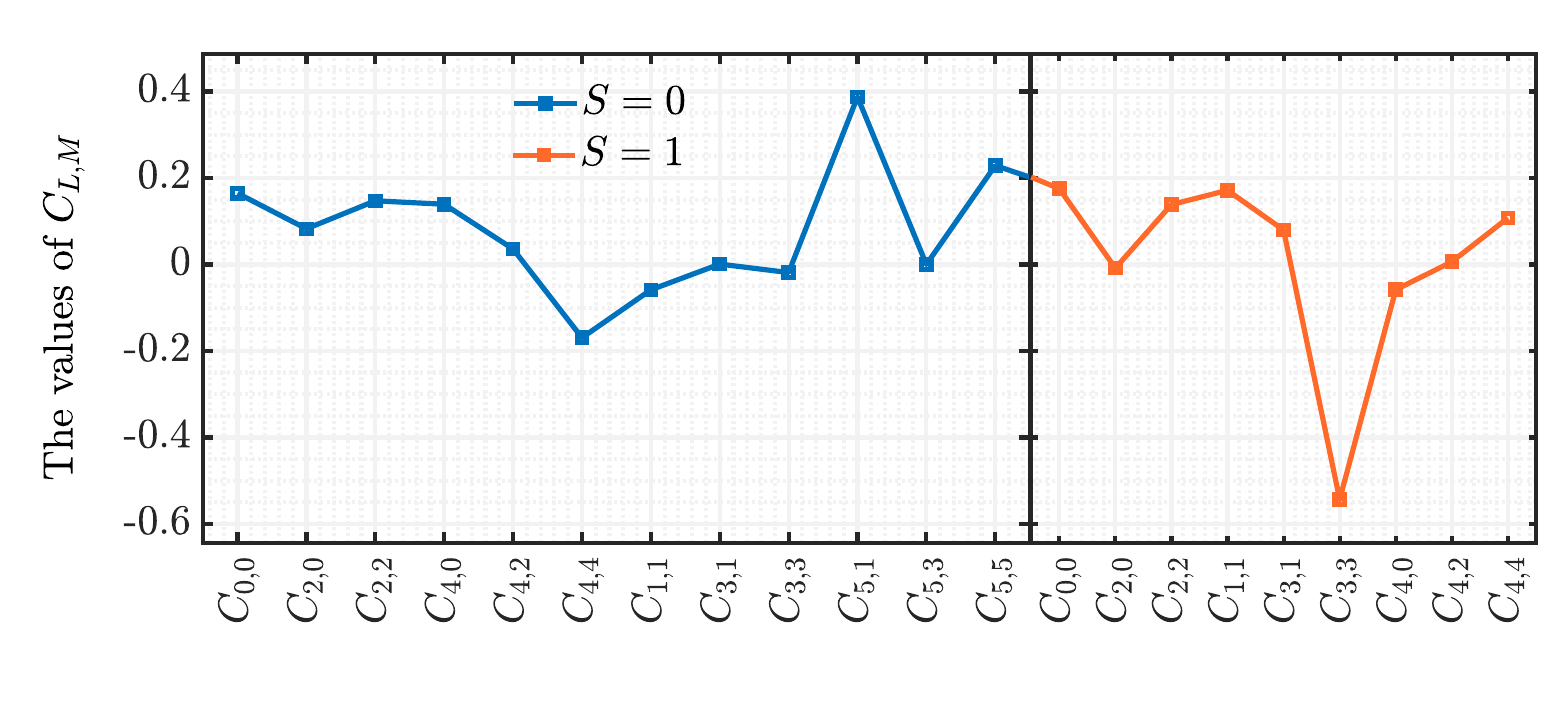}
  \caption{(color online). The expanded coefficents $C_{L,M}$ in Eq.~\eqref{Eq:4} for $\delta = 0$. The left panael corresponds to $S = 0$ and right pannel corresponds to $S = 1$.}
  \label{Fig:4}
\end{figure}

Now, we will first take linear polarization as an example to analyze the characteristics of angular momentum in momentum spectra, the other polarization cases are similar.
The expansion coefficients $C_{L,M}$ for $\delta = 0$ are shown in Fig.~\ref{Fig:4} where the blue line represents the coefficents of spin antiparallel contribution and the orange line represents the spin parallel.

From Fig.~\ref{Fig:4} we can clearly see that the contribution of $Y_{5,1}\sim \sin \vartheta (21 \cos^4\vartheta -14\cos^2\vartheta +1)$ is the largest in the spin antiparallel states.
The contribution of the $Y_{3,3}\sim \sin^3 \vartheta$ partial wave in the spin parallel alignments is the
highest among all the spin states, while the contribution of $Y_{1,1}\sim\sin \vartheta$ and $Y_{0,0}$ (constant) are the next highest for $S = 1$.
The average OAM of the created particles can be approximately estimated using the coefficients $\langle L\rangle =  \sum_{LM}\vert C_{L,M} \vert^2 L/(\sum_{LM}\vert C_{L,M} \vert^2) \approx   3$ that
is consistent with $Y_{3,3}$ contributing is largest for $S = 1$ in Fig.~\ref{Fig:4}, which shows that near the symmetry axis, the spin-parallel contribution is higher than the spin-antiparallel contribution, see Fig.~\ref{Fig:3}(a).
When moving away from the symmetry axis, the contribution of the sinusoidal function related to the partial wave is weakened.
The contribution of the partial wave with small $M$ increases that is the $Y_{2,0}$ and $Y_{4,0}$ in the spin states of $S =0 $ are dominated for the particle creation, see Fig.~\ref{Fig:4}.
This makes it understandable why the $S = 0$ and $S = 1$ contributions when moving away from the symmetry axis are comparable in the $5$-photon ring, see Fig.~\ref{Fig:3}(a).

Similarly, for the elliptically polarized case the particle has an average orbital angular momentum $\langle L\rangle = 2$ and the $Y_{2,2}$ wave in the states of $S = 1$ is the dominant contribution.
However, the $Y_{5,5}$ in the states of $S = 0$ has the highest contribution among all spin states, along with the secondary contributions of $Y_{2,0}$, $Y_{4,0}$ and $Y_{5,1}$
waves with comparable contributions to $Y_{5,5}$. This ultimately leads to the contribution of the $Y_{2,2}$ component wave in the states of $S = 1$ being comparable to that of the main component waves $Y_{2,0}$, $Y_{4,0}$, $Y_{5,1}$ and $Y_{5,5}$ in the spin-antiparallel state,
which makes it understandable that the contribution of $S = 1$ is slightly higher than $S = 0$ at the symmetry axis $\vartheta = \pi/2$, see Fig.~\ref{Fig:3}(b).

For circular polarization $\delta = 1$, all photons have the same helicity, and it will impose strong constraints,
leading to partial waves for pairs with $S = 0$ only including $Y_{5,5}$ and those with $S = 1$ including $Y_{4,4}$~\cite{Kohlfurst:2018kxg}.
As a result, the production of particles has the highest ejection probability towards $\vartheta = \pi/2$, as shown in Fig.~\ref{Fig:3}(c).

We also conclude that the number of quantum states with different $M$ is $9$ and $12$ for the $4$- and $5$-photon rings in non-circular polarization,
and quantum states with different spin alignments are $4$.
The ratio of the total number of quantum states to the number of valleys/peaks is exactly an integer multiple of the Keldysh $\gamma_\omega = m\omega/(eE_0)$~\cite{Keldysh:1965ojf}.
Taking the $5$-photon absorption as an example, the total number of quantum states is $N_s = 48$ and the total number of valleys or peaks is $N_v = 8$, which gives $R = N_s/N_v = 2\gamma_\omega$.
In Ref.~\cite{Li_2015}, the value of $R/\gamma_\omega$ is exactly $10$ for the first two rings.
This shows that the distribution of valleys and/or peaks in multiphoton rings is directly related to the $\gamma_\omega$ or frequency $\omega$ for the fixed field strengths.

\section{Summary}\label{summary}

In summary, we investigate the momentum spectra of multiphoton pair production in polarized homogeneous fields by numerical DHW formalism and theoretical PWA.
The distribution of the valleys/peaks in the momentum spectra depends on the degree of polarization of the field.
Meanwhile, we can also identify the positions of the valleys and peaks according to the angular distribution easily.

The $CP$ conservation, including $P$ invariance under electric dipole transition, can lead to the absence of the $S=0$ states.
However, by extracting the $S = 1$ contribution from the partial wave expansion, we can give a good description of the valleys/peaks positions for the $4$-photon ring under non-circular polarization. Meanwhile, for the $5$-photon ring under non-circular polarization, just a few valleys or peaks close to $\pi/2$ can be identified.
In the case of circular polarization, the valleys and peaks structure in the $n$-photon ring can be described more explicit concisely under $CP$ conservation.

We obtain angular distributions for different spin states by incorporating spin antiparallel contributions in the partial wave analysis when the $P$ conservation is discarded.
We find that the spin-parallel contribution is always higher than the spin-antiparallel contribution in the direction perpendicular to the laser propagation.
In addition, we determine the average OAM of the $5$-photon absorption using partial-wave analysis. For $S = 1$, there is a component in the partial wave expansions where $L = M$, and the OAM of this partial wave equals the average OAM.
The average OAM, as the optimal observable, leads to a maximum contribution at $\pi/2$ for $S=1$.

We believe that this PWA approach revealing the valleys/peaks structures of the multiphoton absorption rings can be extended to other fields.
In particular, the valleys as almost vanishing created particles are associated strongly with the possible existence of vortex
structures, which is crucial for understanding the particle angular momentum inherited from photons.

\begin{acknowledgments}

We are grateful to O. Amat and NZ Chen for helpful discussions. This work was supported by the National Natural Science Foundation of China (NSFC) under Grant No. 12375240, No. 11935008
and No. 12104428. The computation was carried out at the HSCC of the Beijing Normal University.
\end{acknowledgments}

\appendix

\end{document}